\documentstyle[prl,aps,epsfig,multicol]{revtex}
\begin{document}
\draft
\title{Percolative phase separation induced by nonuniformly distributed excess oxygens}
\author{Ilryong Kim, Joonghoe Dho, and Soonchil Lee}
\address{Department of Physics, Korea Advanced Institute of Science and %
Technology, Taejon 305-701, Korea}
\date{\today}
\maketitle

\begin{abstract}
The zero-field $^{139}$La and $^{55}$Mn nuclear magnetic
resonances were studied in $\rm La_{0.8}Ca_{0.2}MnO_{3+\delta}$
with different oxygen stoichiometry $\delta$. The signal
intensity, peak frequency and  line broadening of the $^{139}$La
NMR spectrum show that excess oxygens have a tendency to
concentrate and establish local ferromagnetic ordering around
themselves. These connect the previously existed ferromagnetic
clusters embedded in the antiferromagnetic host, resulting in
percolative conduction paths. This phase separation is not a
charge segregation type, but a electroneutral type. The
magnetoresistance peak at the temperature where percolative paths
start to form provides a direct evidence that phase separation is
one source of colossal magnetoresistance effect.
\end{abstract}


\begin{multicols}{2}
Since the discovery of colossal magnetoresistance (CMR) effect in
$\rm La_{1-x}Ca_{x}MnO_3$ (LCMO)\cite{jin}, many theoretical and
experimental works have been done to find the physical mechanism
of CMR effect because of their interesting physical properties and
application potential. The first explanation of the most
interesting physical property of LCMO, the simultaneous occurrence
of the paramagnetic to ferromagnetic and insulator to metal
transitions, was the simple double exchange model given by Zener
in 1951\cite{zener}. However, Millis pointed out that the
resistivity of Sr-doped manganites cannot be fully explained by
double exchange alone in 1995\cite{millis}. Thereafter, several
theories have been proposed to describe the physical properties of
CMR materials more completely, one of which is phase
separations(PS).

In the low doping range ($x \le 0.2$), the magnetic phase of LCMO
is not homogeneous. The existence of magnetic PS was verified by
the simultaneous observation of ferromagnetic and
antiferromagnetic nuclear magnetic resonance (NMR) signals at low
temperature\cite{allodi}. Since the ferromagnetic metallic regions
are embedded in the antiferromagnetic insulating regions, LCMO in
this range shows ferromagnetic insulating behavior
macroscopically. Since then, PS has been suspected as one of the
possible mechanisms of CMR effect. On the other hand, PS has been
also observed near phase transition temperature in LCMO for $0.2<x
<0.5$, which are homogeneous ferromagnetic metals well below phase
transition temperature\cite{gubkin,dho1}. These two kinds of PS
are thought to be originated from different mechanisms, because
the PS observed in the low doping range is the ground state and
stable in a wide temperature range, while the PS observed for
$0.2<x<0.4$ occurs only near phase transition temperature. In this
report, we will focus our discussion on the PS in low doped LCMO.

Thoery predicts two different types of PS in low doped LCMO. One
is the charge segregation type and the other is the electroneutral
type. Yunoki\cite{yunoki} studied the 2-orbital Kondo model
including the classical Jahn-Teller phonons and found that PS is
induced by the orbital degrees of freedom. In such a case, the
charge density of $\rm e_g$ electron is not stable at a special
value of chemical potential, resulting in two regions with
different charge densities. The size of  both regions is expected
to be very small, about the order of nanometer, due to the
extended Coulomb interaction. Recently, Uehara presented TEM
images of $\rm La_{5/8-y}Pr_yCa_{3/8}MnO_3$\cite{uehara} which
shows the mixture of the charge ordering phase of $\rm
La_{0.5}Ca_{0.5}MnO_3$ type and ferromagnetic phase at low
temperature. The sizes of both regions are about 0.5 $\mu m$,
which is too large to be explained by the charge segregation type
PS. On the other hand, Nagaev paid attention to the PS induced by
nonuniformly distributed oxygens\cite{nagaev}. He pointed out that
the regions enriched with oxygen have an enhanced hole density and
the holes establish local ferromagnetic ordering, whereas the
remained regions are poorly conductive and antiferromagnetic due
to the electron-hole recombination. In this case, PS is the
electroneutal type because the densities of holes and excess
oxygens are same in a given region, and the region sizes can be
much larger than those of the charge segregation type PS. Several
reports have supported the existence of PS, but it still remains
unclear which scenario is more correct in low doped LCMO. There
have been many works which showed that oxygen plays an important
role in determining electromagnetic properties of
LCMO\cite{tamura,ju}, but it's effects on PS have never been
studied. In this work, we report that the electroneutral
percolative PS is formed in low doped LCMO by excess oxygens.
Experimental results provide an evidence for the fact that the PS
in the low doping range is one source of CMR.

Two polycrystalline samples of $\rm
La_{0.8}Ca_{0.2}MnO_{3+\delta}$ with different $\delta$ values
were synthesized by the conventional solid state reaction method.
The starting materials were La$_2$O$_3$, MnCO$_3$, and CaCO$_3$.
Calcining and sintering with intermediate regrinding were repeated
in the temperature range of 1000 $^0$C - 1350 $^0$C for four days.
Sample 1 was obtained by annealing in air at 1100 $^0$C for two
days, and sample 2 was obtained by additional grinding, sintering
and annealing of a part of sample 1 in oxygen flow (200 cc/min) at
the same temperatures with sample 1. The crystal structures were
examined by x-ray powder diffraction with Cu $\it K\alpha$
radiation. Both samples were single phase and orthorhombic. The
lattice parameters of sample 1 and 2 were $ a_1$=5.489 \AA, $
b_1$=5.496 \AA, $ c_1$=7.765 \AA, and $ a_2 $=5.502 \AA, $
b_2$=5.507 \AA, $ c_2$=7.777 \AA, respectively. Resistivity was
measured using the conventional four-probe method, and
magnetization was measured by a commercial SQUID magnetometer.
Zero field NMR spectra were obtained by using a spin-echo
technique.

Fig. 1 shows the temperature dependence of magnetization at 1
Tesla. The paramagnetic Curie temperatures (T$\rm _C$) of sample 1
and 2 are 191.5 K and 192.2 K, respectively. The difference of
T$_C$ values less than 1 K indicates that the difference between
$\delta$ values of sample 1 and 2 is less than 0.02\cite{tamura}.
The magnetizations of sample 1 and 2 at low temperature were
almost same, and the magnetic field dependences of the
magnetizations were almost same either.

Though the macroscopic magnetic properties of two samples are very
similar quantitatively, local magnetic environments are quite
different as seen in the $^{139}$La NMR spectra obtained at 78 K
(Fig. 2). In the figure, two differences are noticeable between
the spectra of sample 1 and 2. First, the signal intensity of
sample 2 is about five times that of sample 1, and second, the
resonance frequency of sample 2 is higher and the linewidth is
much broader, especially in the high frequency side. We discuss
about the difference of signal intensity first.

The NMR signal intensity of a ferromagnet in zero field is
proportional to $\eta V H_L /T$, where $\eta$ is the enhancement
factor, V is the volume of the ferromagnetic region of a sample,
and $H_L$ is the local field at the nuclei of interest. Since
magnetization which is proportional to V does not change, the NMR
signal intensity change at a given temperature and a frequency by
extra oxygens is due to the change of the enhancement factor. Fig.
3 provides experimental evidences for this claim. Fig. 3(a) shows
that sample 2 gives the maximal signal at the much lower rf power
than sample 1, meaning that rf field is more enhanced in sample 2
than in sample 1. The rf enhancement factors of sample 1 and 2 are
about 22 and 105, respectively.

Fig. 3(b) shows the normalized NMR signal intensity vs. external
magnetic field obtained at the fixed rf power which makes the
maximal signal in zero field. In this figure, we notice that the
signal of sample 2 decays almost to zero while that of sample 1
decays slowly approaching the saturation field, about 3 koe. These
are the typical responses of single and multi-domain ferromagnets,
respectively. The signal of sample 1 decays a little because the
enhancement factor decreases with external field. On the other
hand, the drastic signal decay of sample 2 means that domain walls
disappear approaching the saturation field. Therefore, these
results show that sample 2 has domain walls while sample 1 does
not. The NMR signal of sample 2 comes mostly from domain walls
because the enhancement factor is usually orders of magnitude
larger in domain walls than in domain in general\cite{turov}. The
size of ferromagnetic clusters embedded in the antiferromagnetic
host of sample 1 is not large enough to form multi-domain state.
The local ferromagnetic orderings generated by excess oxygens
connect some of these ferromagnetic clusters and domain walls are
formed on them.

In fact, these connected ferromagnetic clusters make also
percolative conduction paths as shown in Fig. 4 displaying the
temperature dependence of resistivity. The resistivity of sample 1
shows an insulating behavior except a small bending near phase
transition, while that of sample 2 shows a broad peak in the
temperature range of 170 K - 140 K and a metallic behavior below
140 K. The metallic behavior of sample 2 at low temperature
implies that electric transport paths are formed by excess
oxygens.

The temperature dependence of the La NMR signal intensity of
sample 2 shown in Fig. 5 supports the simultaneous generation of
conduction paths and domain walls. The signal intensity of
homogeneous ferromagnets such as LCMO for $0.2<x<0.5$ well follow
Curie's $T^{-1}$ law except in the narrow region near
T$_C$\cite{dho1}. However, the signal intensity of sample 2
decreases much faster than T$^{-1}$, and almost disappears near
140 K where the metallic behavior fades out (Fig. 4). This means
that the total volume of domain walls decreases as temperature
increases and the ferromagnetic and metallic conduction paths
vanish near 140 K. As approaching this temperature from below,
ferromagnetic clusters are disconnected and therefore conduction
paths are broken continuously.

While the magnetoresistance (MR) curve of stoichiometric
perovskite manganite crystals show only one peak near T$\it _C$,
that of sample 2 shows one more peak near 140 K as seen in Fig. 4.
The MR peak near 170 K is an ordinary CMR peak due to the
suppression of spin fluctuation by external field, while the peak
near 140 K is undoubtedly related with PS. It is worthwhile to
note that the MR near the temperature where the percolative PS is
induced is as large as that near the phase transition temperature.
External field helps connecting clusters somehow.

We now discuss the second difference of the spectra of sample 1
and 2, the difference in resonance frequency and linewidth. The
local field $H_L$ at the position of a non-magnetic La$^{3+}$ ion
can be described as $$ H_{L} = A \sum_{\scriptstyle j} n_j \mu _j
+ H_{d-d}, $$ where $A$ is the transferred hyperfine coupling
constant and $n_j$ is the number of the j-site Mn moments $\mu
_j$, surrounding the La ion. $H_{d-d}$ is the dipolar field summed
over all Mn magnetic moments. In perovskite manganites, the
dipolar field is negligible and the main contribution to $H_L$
comes from the transferred hyperfine field. The transferred
hyperfine field is thought to be produced by the $\pi$ type
overlapping between the Mn $t_{2g}$ electron wave function and the
oxygen $|2p_{\pi}\rangle$ wave function, and the $\sigma$ bonding
of the oxygen with the $|sp^3\rangle$ hybrid states of the
La$^{3+}$ ion\cite{papa}. That is, an indirect transferred
hyperfine field of the Fermi contact type is mediated by oxygens.
Therefore, the constant $A$ is a function of the distance between
oxygens and a La$^{3+}$ ion, and the number of the oxygens
surrounding the La$^{3+}$ ion. $n_j$ and $\mu _j$ are almost same
in two samples. The distance between oxygens and a La$^{3+}$ ion
is not an important factor making the difference of the NMR peak
frequencies because the peak frequency of sample 2 is higher than
that of sample 1 even though the lattice constants of sample 2 are
slightly larger than those of sample 1\cite{papa2}. Therefore, the
difference of peak frequencies should be attributed to the
difference of the number of oxygens surrounding a La$^{3+}$ ion.
That is to say, the peak frequency of sample 2 is increased due to
the excess interstitial oxygens. There is a report that
LaMnO$_{3+\delta}$ with excess oxygens is characterized by cation
vacancy in La and Mn sites rather than by interstitial
anions\cite{tofi}. In this case, however, the lattice parameters
decrease as $\delta$ increases contrary to our case\cite{mah}.
Moreover, the NMR spectrum of LaMnO$_{3+\delta}$ having cation
vacancy is well fitted by a single gaussian curve while our sample
2 is not as discussed below.

The La NMR spectrum of sample 2 is asymmetric and broader than
that of sample 1 in the high frequency side. Since $\delta$ is
less than 0.02, homogeneous distribution of oxygens cannot enhance
the signal in the high frequency side as much as the spectrum in
Fig. 2. Therefore, oxygens aggregate to make local ferromagnetic
orderings in consistence with Nagaev's claim that oxygens have a
tendency to concentrate. One of the reasons why conduction paths
are continuously disconnected as approaching 140 K could be the
distribution of excess oxygens becoming more and more uniform as
temperature increases.

Contrary to La nuclei, the local field at Mn nuclei is negligibly
influenced by the local distribution of oxygens because the
hyperfine field of the direct Fermi contact type generated by it's
own 3$d$ electrons is much stronger. Therefore, the gaussian shape
of the Mn NMR spectrum does not change by the presence of excess
oxygens as shown in the inset of Fig. 2. The $^{55}$Mn NMR spectra
are motionally narrowed by the fast hopping of e$_g$ electrons
between Mn$^{3+}$ and Mn$^{4+}$ ion sites\cite{matsumoto,dho2}.
Only the Mn nuclei in ferromagnetic regions with the delocalized
e$_g$ electrons contribute to the $^{55}$Mn NMR signal. The local
field at Mn nuclei is proportional to the number of average
delocalized e$_g$ electrons. The peak frequency shift by excess
oxygens is not less than 10 Mhz. If the distribution of holes is
uniform, such a shift corresponds to $\delta
\sim$0.075\cite{dho2}, while $\delta$ of sample 2 is less than
0.02. This means that the holes are concentrated in ferromagnetic
regions. The results of La and Mn NMR imply the concentrated
oxygens and holes in ferromagnetic regions, respectively. This
support the fact that the PS in low doped LCMO is the
electroneutal type. Moreover, considering the easy formation of
conduction paths by the aggregation of excess oxygens less than
0.7 \%, the size of ferromagnetic clusters are not so small as
predicted by the charge segregation type PS.

In conclusion, the excess oxygens have a tendency to aggregate and
change surroundings into ferromagnetic phase. These local
ferromagnetic regions connect ferromagnetic clusters previously
existed in stoichiometric samples, which are suspected to be also
generated by inhomogeneous distribution of oxygens. This
connection produces percolative conduction paths on which domain
walls are formed. The observed PS is rather a electroneutral type
than a charge segregation type. As temperature increases, the MR
peak was observed at the temperature where the percolative PS
disappears in addition to the ordinary peak near the phase
transition temperature.

\newpage

\begin{figure}[t]
\epsfig{file=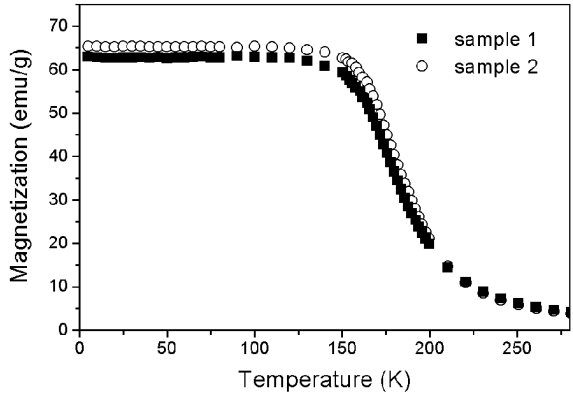, width=8cm} \vspace{0.5em}
\caption{\narrowtext The temperature dependence of magnetization
at 1 Tesla.} \label{fig.1}
\end{figure}

\begin{figure}[t]
\epsfig{file=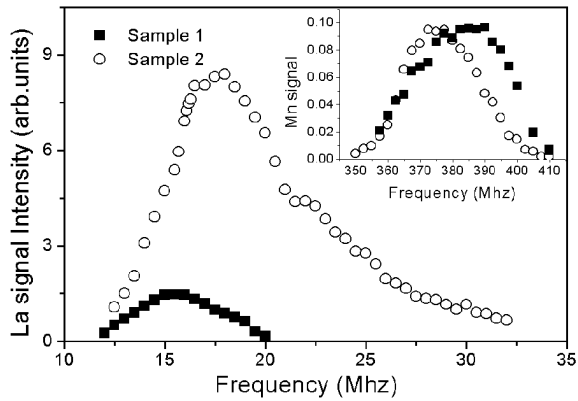, width=8cm} \vspace{0.5em}
\caption{\narrowtext The zero field $^{139}$La NMR spectra
obtained at 78 K. The frequency and and the spin-spin relaxation
time dependences of signal intensity were removed. Inset: the zero
field $^{55}Mn$ NMR spectra obtained at 78 K. The intensity of
sample 1 is about forty time amplified.} \label{fig.2}
\end{figure}

\begin{figure}[t]
\epsfig{file=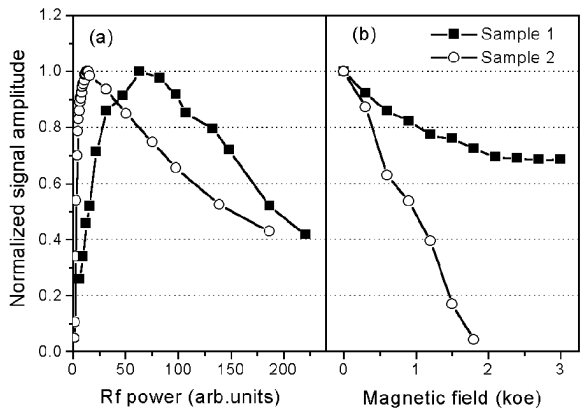, width=8cm} \vspace{0.5em}
\caption{\narrowtext The rf-power dependence(a) and external
magnetic field dependence(b) of $^{139}$La NMR signal intensity
obtained at 78 K.} \label{fig.3}
\end{figure}

\begin{figure}[t]
\epsfig{file=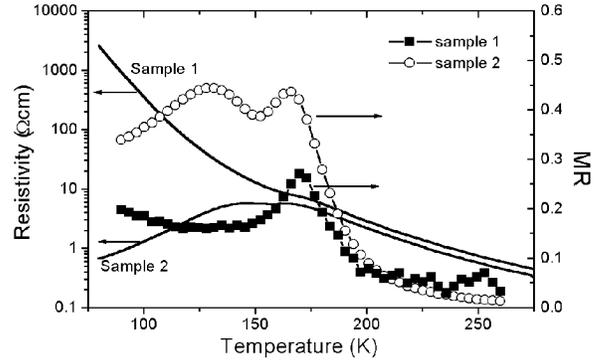,width=8cm} \vspace{0.5em}
\caption{\narrowtext The temperature dependence of resistivity and
magnetoresistance (MR). The MR value is defined as $(\rho(0) -
\rho(9 ~{\rm koe}))/\rho(0)$.} \label{fig.4}
\end{figure}

\begin{figure}[t]
\epsfig{file=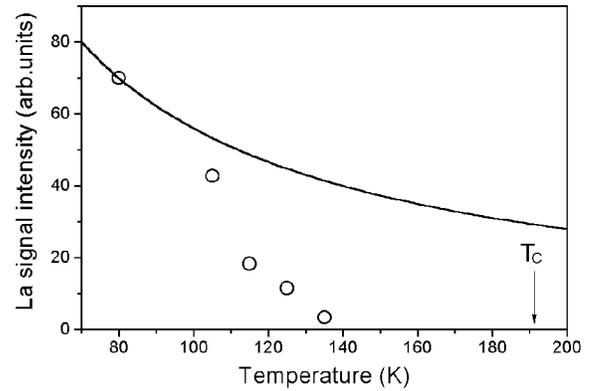,width=8cm} \vspace{0.5em}
\caption{\narrowtext The temperature dependence of the $^{139}$La
NMR intensity of sample 2. The signal dependence on the spin-spin
relaxation time and frequency were carefully eliminated from the
raw data. The solid line represents a T$^{-1}$ curve.}
\label{fig.5}
\end{figure}

\end{multicols}
\end{document}